# Spin-dependent Schottky barriers and vacancy-induced spin-selective Ohmic contacts in magnetic vdW heterostructures

Hongxing Li,[a,b] Yuan-Kai Xu,[a,b] Zi-Peng Cheng,[a,b] Bin-Guang He,[a,b] and Wei-Bing Zhang[*a,b]

a *School of Physics and Electronic Sciences, Changsha University of Science and Technology, Changsha 410114, People's Republic of China*
b *Hunan Provincial Key Laboratory of Flexible Electronic Materials Genome Engineering, Changsha University of Science and Technology, Changsha 410114, People's Republic of China*
Email: zhangwb@csust.edu.cn

**Abstract**
The 2D ferromagnets, such as $CrX_3$ (X=Cl, Br and I), have been attracting extensive attentions since they provide novel platforms to fundamental physics and device applications. Integrating $CrX_3$ with other electrodes and substrates is an essential step to their device realization. Therefore, it is important to understand the interfacial properties between $CrX_3$ and other 2D materials. As an illustrative example, we have investigated the heterostructures between $CrX_3$ and graphene ($CrX_3$/Gr) from first-principles. We find unique Schottky contacts type with strongly spin-dependent barriers in $CrX_3$/Gr. This can be understood by synergistic effects between the exchange splitting of semiconductor band of $CrX_3$ and interlayer charge transfer. The spin-asymmetry of Schottky barriers may result in different tunneling rates of spin-up and down electrons, and then lead to spin-polarized current, namely spin-filter (SF) effect. Moreover, by introducing X vacancy into $CrX_3$/Gr, an Ohmic contact forms in spin-up direction. It may enhance the transport of spin-up electrons, and improve SF effect. Our systematic study reveals the unique interfacial properties of $CrX_3$/Gr, and provides a theoretical view to the understanding and designing of spintronics device based on magnetic vdW heterostructures.

**Introduction**
The intrinsic 2D ferromagnetic materials are the long-sought goals for both the fundamental physics and device application. According to Mermin-Wagner theorem, the long-range magnetic order is destroyed by thermal fluctuation at finite temperature.[1] However, this restriction can be eliminated by magnetic anisotropy. In 2017, the single layer $CrI_3$ was exfoliated successfully and proved to be an intrinsic Ising ferromagnet.[2] The discovery of $CrI_3$ intrigues extensive research on 2D magnets. Since then, a lot of 2D magnets have been discovered, and many novel properties have been revealed. For example, $CrI_3$ was proved to be a bosonic Dirac material,3 and novel electron tunneling phenomenon has been fund in $CrBr_3$.[4]

Magnetic tunnel junction (MTJ) is one of the most important spintronic device. Generally, one form of MTJ is the heterostructure composed by metal/ferromagnet/insulator/ferromagnet/metal, where the ferromagnetic, insulator and metal layer are spin-filter (SF), tunneling barrier and electrode, respectively.[5] By switching the interlayer magnetic alignment, such as from antiparallel to parallel, different tunnel resistances will be obtained, namely magnetoresistance effect. Unfortunately, the number of materials that can be used to design MTJs is limited,[6] and the higher magnetoresistance (MR) is desired at present.[7] The emergence of 2D ferromagnetic materials and magnetic vdW heterostructures stacking by various 2D materials revolutionize the MTJ.[8] Comparing with the traditional MTJs, the atomic flat interface in vdW heterostructures can render uniform tunneling, which may result in higher MR. In addition, the interlayer space can sever as natural tunneling barriers.

The electrode is another key component in the MTJs-based electronic devices. Among various 2D electrode candidates, graphene is one of most widely used material due to its high-performance and low-cost. For example, the graphene has been used in $CrX_3$-based spin device including the multiple-spin-filter MTJs,[4,9-15] and very high large tunneling MR values up to 19 000% has been reported. Furthermore, as vacancy is inevitable in 2D materials, which is known to affect the physical properties of host remarkably. Therefore, it is crucial to understand the properties of interface between $CrX_3$ and graphene, and explore the effect of defect on the properties of $CrX_3$/Gr heterostructures.

In present paper, by first-principles calculations, we systematically study the electronic and interfacial properties of $CrX_3$/Gr vdW heterostructures. We find that the Schottky contacts form at the interface. Distinct from ordinary semiconductor-metal interface, the Schottky barriers $\Phi$ exhibits strong spin-dependence. Specially in the case of $CrI_3$/Gr, the Schottky barriers for electron $\Phi e$ in spin-up direction is 30 meV, but up to 760 meV in spin-down. This interesting phenomenon will lead to different tunneling rates of spin-up and spin-down electrons, and the SF effect. In addition, we find the X vacancy induces interesting defect bands into the spin-up channel of $CrX_3$. Furthermore, in $CrX_3$/Gr with X vacancy ($CrX_3@X_v$), the Fermi level crosses the spin-up band of $CrX_3$ layer, which leads to Ohmic contact in spin-up direction, and possibly enhance the SF effect.

**Results and discussions**

**Single-layer $CrX_3$**
As presented in Figure 1, the Cr atoms in single-layer $CrX_3$ arrange in one plane, forming a honeycomb lattice. Each Cr atom is surrounded by six nearest neighbor X atoms that arranged in an octahedra. Due to the octahedral crystal field, the d orbit splits into low-energy $t_{2g}$ and high-energy $e_g$ orbit. At the same time, the X atoms provide superexchange path between Cr atoms, which are responsible for the long range ferromagnetic order,[16] and even the out-plane anisotropy.[17] The optimized in-plane lattice constants of $CrX_3$ are 6.10 ($CrCl_3$), 6.35 ($CrBr_3$) and 7.01 Å ($CrI_3$), due to the increasing of halogen atomic radii. The lattice constant of graphene is optimized to be 2.45 Å. These results agree well with the reported values.[18-20] In our study, the supercell of CrCl3/Gr, CrBr3/Gr and CrI3/Gr used in calculation are composed by 2×2 CrCl3, 2×2 CrBr3 and $\sqrt{3} \times \sqrt{3}$ CrI3 and 5×5 graphene, with total 74, 74 and 68 atoms, respectively. As the stiffness of graphene is larger than CrX3, we keep graphene unstrained. The lattice mismatches lead to 0.4%, 3.4% and 1.0% strain in CrCl3, CrBr3 and CrI3, respectively.

Figure 2(a)-(c) depict the spin-polarized band structures of single-layer CrX3. We can see that these compounds are half-semiconductors with gaps of 1.58, 1.36 and 1.11 eV for CrCl3, CrBr3 and CrI3, respectively. The decreasing trend of band gap from CrCl3 to CrI3 is in line with the crystal field strength from Cl to I. More interestingly, the band structures also show strong spin polarization. For instance, the conduction band minimum (CBM) of CrBr3 in spin-up and spin-down direction are about 1.09 eV and 1.63 eV. We further calculate the band structures of defective CrX3. The defective CrX3 are modeled by removing a X atom from lattice matched CrX3/Gr heterostructures. The corresponding atomic vacancy concentrations are 4.2%, 4.2% and 5.6% for CrCl3, CrBr3 and CrI3, respectively. The band structures are shown in 2(d)-(f). Comparing with pristine CrX3 layer, the most obvious difference is the emergence of defect bands in gap region, which are resulted from the dangling bonds of Cr atoms that near to X vacancy.[21,22] Interestingly, all the defect bands distribute in spin-up channel, which indicate a strong spin selectivity, and reduced the gap to 0.46, 0.35 and 0.15 eV for CrCl3, CrBr3 and CrI3, respectively. This is different from the nonmagnetic 2D semiconductor MoS2, in which the defect states induced by S vacancy is not spin-polarized.[23]

**Perfect CrX3/Gr**

The atomic structures of CrI3/Gr and CrCl3/Gr (CrBr3/Gr) heterostructures are sketched in Figure 3(a) and (b). Firstly, we try to determine the most stable stacking configuration by comparing the total energies of various heterostructures obtained by sliding CrX3 layer along zigzag and armchair directions. However, the total energies difference is tiny. Hence the interactions between CrX3 and graphene do not exhibit evident site selectivity, which is in line with reports about of CrI3/Gr, CrBr3/Gr and RuCl3/Gr heterostructures.[19,20,24] Nevertheless, in our previous study, we found the strong site-dependence of total energies in CrX3/silicene and CrX3/germanene heterostructures.[25] The difference may be originated from the bulking of silicene and germanene. The interlayer distances d are measured to be about 3.36, 3.45 and 3.47 Å for CrCl3/Gr, CrBr3/Gr and CrI3/Gr, respectively. These distances are much large than the sum of covalent atomic radii of C and X, indicating there are no chemical bonds between C and X atoms, and the interlayer interactions are dominated by van der Waals force. However, the interlayer distances are larger than the distances in CrX3/metal heterostructures.[26] To evaluate the interlayer interaction strength, we calculated the interlayer binding energy $E_b$ by

$$E_b = (E_{Gr} + E_{CrX3} - E_{tot})/A \qquad (1)$$

where $E_{Gr}$, $E_{CrX3}$ and $E_{tot}$ are the total energies of isolated graphene, CrX3 and CrX3/Gr heterostructures, respectively. A is the area of the interface. The calculated $E_b$ for CrCl3/Gr, CrBr3/Gr and CrI3/Gr are 9.68, 17.38 and 14.34 meV/Å$^2$, respectively. The results are consistent with previous studies,[19,20] also comparable to the 18.4 meV/Å$^2$ of GaSe/Gr heterostructures.[27] However, the binding energies between CrX3 and graphene are much smaller than the binding energies, about 1 J/m$^2$ (62.5 meV/Å$^2$), between CrX3 and metal surfaces at the same level.[26] The very different interlayer binding energies in CrX3/Gr and CrX3/metal heterostructures is easy to understand. The graphene is dangling-bond free, while redundant electrons may be hosted by metal surface, which may participate in the bonds between CrX3 and the metal surface.

To further evaluate the interlayer interaction, we calculate the charge density difference (CDD) of CrX3/Gr heterostructures, which is defined by

$$\Delta\rho = \rho(heterostructure) - \rho(graphene) - \rho(CrX3) \qquad (2)$$

Similar features can be found in the results, so we representatively show the result of CrBr3/Gr in Fig 3(c) and (d). As we can see, the charge depletion occurs at the graphene layer with charge accumulation at Br atoms of CrBr3 positioned close to graphene, and there is no charge redistribution at the outside Br atoms. Interestingly, we also observe a small amount of charge accumulation at Cr atoms. This result indicates graphene is a charge-donating substrate for CrX3, and it is expectable as the calculated work function of graphene (4.28 eV) is smaller than the electron affinity of CrX3 (5.38, 4.73, and 4.69 eV for CrCl3, CrBr3 and CrI3, respectively). This is different to the MoS2/Gr heterostructure, in which the charge transfers from MoS2 to graphene.[28] The interlayer charge transfer can give rise to a built-in electric field points from graphene layer to CrX3. Our result identifies with experimental result that the CrI3 is n-doped when contact with graphene.[29] Moreover, previous studies have proved that charge doping can effectively tune the magnetic properties of CrX3.[25,30,31] Therefore, the interlayer charge transfer may be one way to tune the magnetic properties of CrX3 in CrX3/Gr heterostructures. Besides, in the experimental study of Mak et. al,[32] the observed magnetic circular dichroism signal of BN/CrI3/Gr and BN/CrI3/BN systems show significant difference. For comparison, we also study the CrI3/BN heterostructure, and find the charge transfer between CrI3 and BN layer is much less than that between CrI3 and graphene. Therefore, the different magnetic properties of BN/CrI3/Gr and BN/CrI3/BN systems may be originated from the different charge transfer and built-in electric field in CrI3/Gr and CrI3/BN interfaces. Afterwards, we quantificationally calculate the transferred charge Δq per CrX3 unit by Bader charge analysis.[33] The results are 0.072, 0.015 and 0.003 e for CrCl3/Gr, CrBr3/Gr and CrI3/Gr, which are in line with the order of electron affinities of CrX3.

The projected density of states (PDOS) are calculated to reveal the electronic properties of CrX3/Gr, and we present the PDOS of CrBr3/Gr in Fig 4. In single-layer CrX3, the d orbits of Cr atom split into triple degenerate $t_{2g}$ orbit and double degenerate $e_g$ orbit. However, the PDOS demonstrates the partial breaking of degeneracy of d states in CrX3/Gr. For example, the $d_{z^2}$ orbit is no longer degenerate with $d_{x^2-y^2}$ orbit. This is due the broken inversion symmetry and formation of z-direction built-in electric field.

Therefore, the vdW engineering provides an effective way to tune the orbital characteristic of 2D materials. However, the orbits of CrX3 retain partial degeneracy, such as the $d_{xy}$ and $d_{xz}$ orbits, and different from the complete breaking in CrX3/metal heterostructures.[26] By carefully structural checking, we find the atomic structure of CrX3 in heterostructure is almost the same as single-layer, which is in contrast with great deformation when CrX3 interacts with metal surface.[26] Fig 4(d) depicts the PDOS of graphene. The states near Fermi level are disturbed to some extent, indicating weak hybridization between graphene and CrX3. In addition, the states of spin-up and spin-down direction are no longer symmetric. Therefore, net magnetic moments are induced in graphene due to the magnetic proximity effect, which will break the time reversal symmetry in graphene, and may bring about quantum phase transition.[19,20]

Next, we calculate the spin-polarized band structures of the CrX3/Gr heterostructures. The band structures of these three systems show similar features, so we show the band of CrBr3/Gr in Fig 5. We can see that the Dirac cone of graphene is well preserved, confirming the vdW interaction. It is similar to the graphene on nonmagnetic semiconductor GaSe and MoS2.[27,28] However, the Dirac cone moves upward and submerge into the conduction band, consisting with the result that charge transfers from graphene to CrX3. Our results are very similar to the reports,[19,20] but different from Klein's result,[10] in which there is a certain amount of states distributed across Fermi level. This difference possibly comes from the different layers of graphene and CrI3 in supercell that used in calculation. To visualize the real-space distribution of bottom of conduction bands in spin-up and spin-down direction, we plot the partial charge density. Due to the dense and flat bands, the charge densities of four bands at the bottom of conduction band in each spin direction are evaluated. The results are shown in Fig 5(d) and (e). We can be informed that the charge mainly distributes around the Cr atoms in both spin directions. It indicates the transport process in both spin directions is principally involved by the d states of Cr atoms. However, the spin-down partial charge density extents widely than spin-up, and there is more charge distributes around X atoms. Interestingly, there is a small quantity of $p_z$ orbital charge distributed at graphene.

To reveal the energy level alignment, and illustrate the effects of interlayer interactions on band structures of CrX3, we project the band of heterostructures to CrX3 layer in spin-up and spin-down direction. Fig 5(b) and (c) depict the projected bands of CrBr3 layer. We find the band structures of CrX3 in heterostructure are very similar to single-layer. For example, the gaps in spin-up and spin-down direction of CrBr3 layer in heterostructure are 1.34 and 2.42 eV, which are very close to the 1.36 and 2.57 eV of single-layer CrBr3. Therefore, the electronic properties of CrX3 are influenced slightly by the interlayer interactions. This is similar to the case of GaSe/Gr heterostructure, where the change of GaSe layer gap is only about 0.01 eV by the interlayer interaction.[27]

Now we turn to the contact properties of CrX3/Gr, as the electronic transport through the interface between CrX3 and graphene is an interesting topic now. When electron transports through metal-semiconductor interface, the barrier it encountered may be composed by two parts, namely tunneling barrier ($\Phi_{TB}$) and Schottky barrier ($\Phi$).34,35 In vdW metal-semiconductor interface, because the the large vdW gap and weak interlayer orbital hybridization, the $\Phi_{TB}$ may be pronounced. To evaluate the tunneling barrier height, we calculate the effective electrostatic potential along Z-direction ($V_{eff}$). Fig 6(a) depict the $V_{eff}$ of CrBr3/Gr, and illustrates the definition of $\Phi_{TB}$, which is the potential difference between the vdW gap and CrX3 layer. The calculated $\Phi_{TB}$ are 7.0, 6.5 and 5.4 eV for CrX3/Gr (X=Cl, Br and I). The values are larger than the case of MoS2/metal interfaces, in which $\Phi_{TB}$ is smaller than 1 eV.[36]

Afterwards, we try to evaluate the Schottky barrier ($\Phi$). As experiments mainly probe the electrons tunneling process, we merely consider the Schottky barrier height for electron, $\Phi_e$. Interestingly, due to the exchange splitting in magnetic semiconductor, the electrons with different spin direction will exhibit different transport properties. Therefore, according to Schottky-Mott rule,[37] the barrier heights for spin-up/down electrons $\Phi_{e\_up/dn}$ can be defined as

$$\Phi_{e\_up/dn} = E_{CBM\_up/dn} - E_F \qquad (3)$$

where $E_F$ is the Fermi energy, $E_{CBM\_up/dn}$ is the energy of conduction band minimum (CBM) in spin-up/down direction of CrX3 in the heterostructure. The measured values are listed in Table 2. The values of $\Phi_e$ for spin-up are much less than for spin-down. For instance, in CrBr3/Gr, the $\Phi_{e\_up}$ is 61 meV, while $\Phi_{e\_dn}$ is as large as 547 meV. As the tunneling barrier is Coulomb interaction nature and does not exhibit spin selectively, the different $\Phi_{e\_up}$ and $\Phi_{e\_dn}$ signify different barriers will be encountered by spin-up and spin-down electrons when they transport through the interface. The spin-dependent barriers will give rise to different tunneling rates for spin-up and spin-down electrons, thus generate spin-filter effect. This is consistent with the remarkable spin-filter effect observed by experiments. [10,11,29] Kim et. al[14,15] have studied the electronic transport barrier based on devices composed by bilayer CrX3 and few layers graphene, they found the barrier height is spin-dependent, such as CrBr3, the height is 477 meV and 599 meV for spin-up and down-electron. Our result is qualitatively consistent with the report. However, big discrepancy exits among the values. There are several reasons may account for the divergence. Firstly, under different bias and temperature, electrons transport across the metal-semiconductor may be dominated by different mechanisms.[38] The barrier height extracted by Kim et. al is based on Fowler-Nordheim tunneling under high bias at low temperature 1.4 K. However, the Schottky barrier play a dominant role in Schottky emission under low bias and higher temperature.[39] Secondly, the GGA-PBE functional is well-known for imprecise estimation of band gap. Thirdly, the different material layers used in experiment and our calculation.

**The introduction of X vacancy**

We further study the properties of CrX3/Gr with one X vacancy, denoted by CrX3/Gr@Xv. Structurally, the X vacancy can be distributed inside ($V_{in}$) and outside ($V_{out}$) in heterostructure, as shown in Figure 7(a). To determine the stable configuration, we calculate the binding energy $E_b$ for $V_{in}$ and $V_{out}$ configurations, as shown in Figure 7(b). In all these hybrid systems, the $E_b$ for $V_{out}$ is larger than $V_{in}$, suggesting the X vacancies are inclined to distribute at $V_{out}$ site in these heterostructures. However, the X vacancies show slight effects on $E_b$. In CrCl3/Gr, the Cl vacancy enhances the interlayer binding slightly for both $V_{in}$ and $V_{out}$, similar to the enhancement of interlayer binding caused by intrinsic atomic vacancies in BN/Gr and MoSe2/metal heterostructures.[40,41] While in CrBr3/Gr and CrI3/Gr, the interlayer binding will be weaken by Br(I) vacancy, especially the vacancy distributed at $V_{in}$. The X vacancy has competitive effects on the interlayer binding. On one hand, the vacancy results in dangling bonds, which can increase the interlayer charge transfer, and lead to positive contribution to $E_b$. On the other hand, X atom is the acting point of vdW interaction, thus the missing of X will reduce the interlayer binding. Then, as the distance between graphene and inside X atom is smaller than outside X atom, the missing of inside X atom has more negative influence on interlayer interaction than outside X atom. Therefore, the $E_b$ for $V_{out}$ is larger than $V_{in}$. Furthermore, as the vdW interaction is positive correlated with atomic numbers, the missing of heavier atom leads to greater loss of vdW interaction. As competitive results, the Cl vacancy increases $E_b$, while Br (I) decreases $E_b$.

Next, we calculate the band structures of the CrX3/Gr with atomic vacancy. Despite the different atomic structure and interlayer binding energy for $V_{in}$ and $V_{out}$ configurations, the band structures seem to be structural insensitive. Therefore, we present the band structures of the systems at the energetically favorable $V_{out}$ configuration in Fig 8. The Dirac cone of graphene is well preserved, the same as the perfect heterostructures. However, we can observe obvious interactions between the defect bands and $\pi$ band of graphene. Such as CrI3/Gr, the defect bands are fractured markedly by interaction.

Finally, we plot the projected band of CrX3 layer in spin-up and down direction, as shown in Fig 8. The spin-down bands are very similar to the bands of perfect heterostructures, except for the positions of band edges. For example, the band edges of CrI3 layer in spin-down direction are -1.12 and 1.05 eV, moving up about 0.2 eV relative to the -1.35 eV and 0.76 eV in perfect CrI3/Gr. The difference in band edges can mainly be attributed to the n-doping by X vacancy. More interestingly, in spin-up direction, the Fermi level crosses the spin-up band of CrX3 layer, it suggests the Ohmic contact in spin-up direction in CrX3/Gr. The formation of spin-up direction Ohmic contact is due to the spin-selective defect band induced by X vacancy and interlayer charge transfer. It will enhance the transport of spin-up electrons, and then the SF effect, and may even inject current to graphene with 100% spin-polarization. In addition, the tunneling barriers $\Phi_{TB}$ in CrX3/Gr@Xv are 6.5, 6.09 and 5.08 eV, reduced by about 0.4 eV comparing to the perfect ones. Hence, the X vacancy can entirely enhance conductivity of the interface.

**Conclusion**

In summary, a systematic study has been conducted to the CrX3/Gr vdW heterostructures. The results show the charge transfer from graphene and leading to n-doping in CrX3. Due to the unique band structures of CrX3, the Schottky barriers in heterostructures for electrons $\Phi_e$ demonstrate strong spin-dependence, which may result in SF effect and be responsible for the observed large TMR values. In addition, upon the introduction of X vacancy, spin-selective Ohmic contacts form in CrX3/Gr with the reduced tunneling barrier. Therefore, the defect-engineering may be an effective way to tune the spintronics properties of magnetic vdW heterostructures.

**Calculation method**

All our spin-polarized density functional calculations are carried out using the Vienna ab initio simulation package (VASP).[42,43] The ion-electron interactions are described by projected augmented wave (PAW) method with cutoff energy of 500 eV.[44] The generalized gradient approximation is used to evaluate the exchange-correction.[45] DFT-D2 method proposed by Grimme[46] is adopted to account for the interlayer van der Waals interactions. A vacuum thickness of 18 Å is used to avoid the artificial interaction between images of slabs. The convergent criterion of electronic step is set to $1 \times 10^{-5}$ eV and the ions will be relaxed until the forces acted on each ion is smaller than $1 \times 10^{-2}$ eV/Å. Brillouin zone sampling is done by a 5×5×1 Γ-center mesh.[47]

**Acknowledgement**

This work was supported by National Natural Science Foundation of China (Grant No.11847157 and No.11874092) the Fok Ying-Tong Education Foundation, China (Grant No. 161005), the Planned Science and Technology Project of Hunan Province (Grant No. 2017RS3034), Hunan Provincial Natural Science Foundation of China (Grant No. 2016JJ2001 and 2019JJ50636), and Scientific Research Fund of Hunan Provincial Education Department (Grant No. 16B002 and 18C0227)

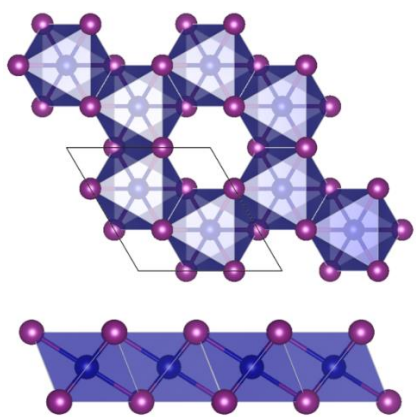

Figure 1 The top and side view of the atomic structure CrX3. The blue and purple balls represent the Cr and X atoms.

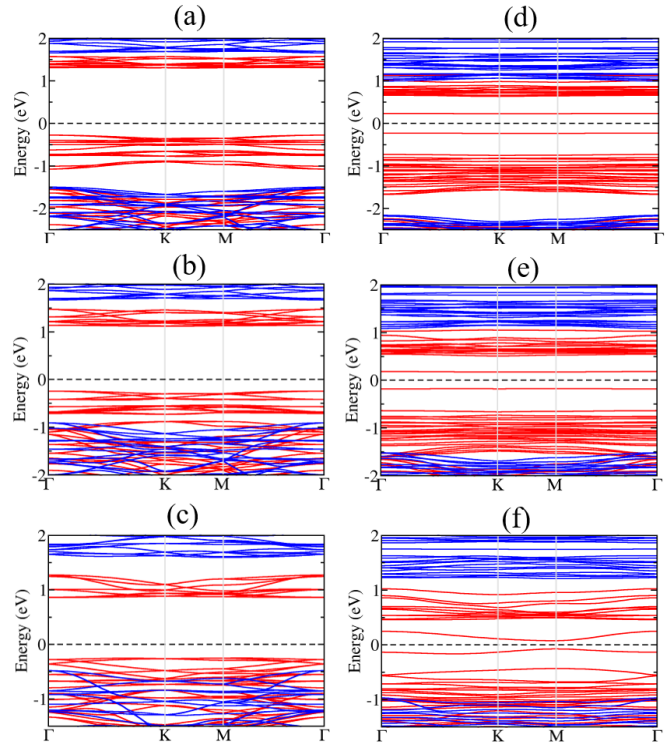

Figure 2 (a)(b) and (c) the spin-polarized band structure of pristine single-layer CrCl3, CrBr3 and CrI3. (d)(e) and (f) the spin-polarized band structure of defective single-layer CrCl3, CrBr3 and CrI3.

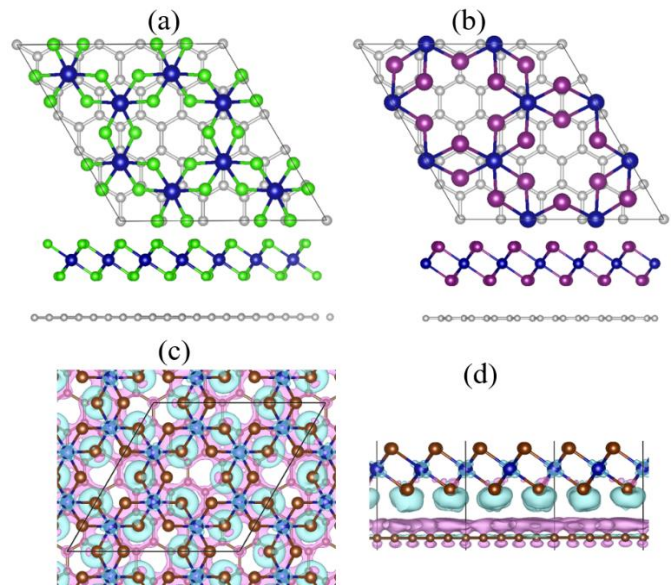

Figure 3 (a)(b) The atomic structure sketch of CrCl3/Gr (CrBr3/Gr) and CrI3/Gr heterostructures. (c)(d) The top and side view of the charge density difference (CDD) of CrBr3/Gr. The purple and blue denote the charge depletion and accumulation, and the isosurface is set to 0.0001 e/Å$^3$

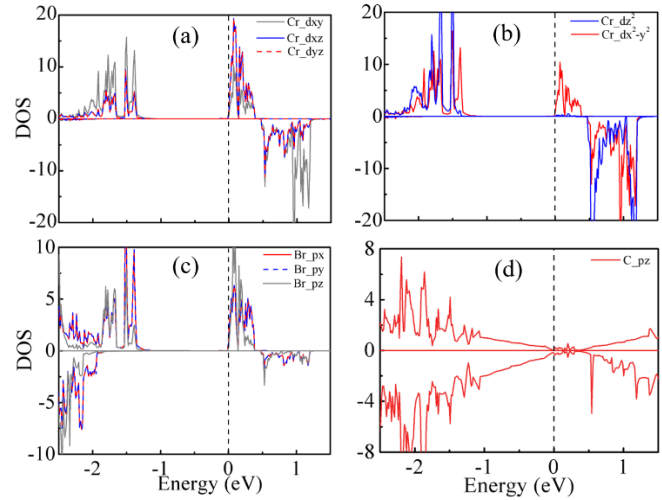

Figure 4 The projected density of states (PDOS) of (a)(b) the d orbits of Cr atom, (c) the p orbits of Br atom, (d) p$z$ orbit of C atom. The Fermi level is set to zero.

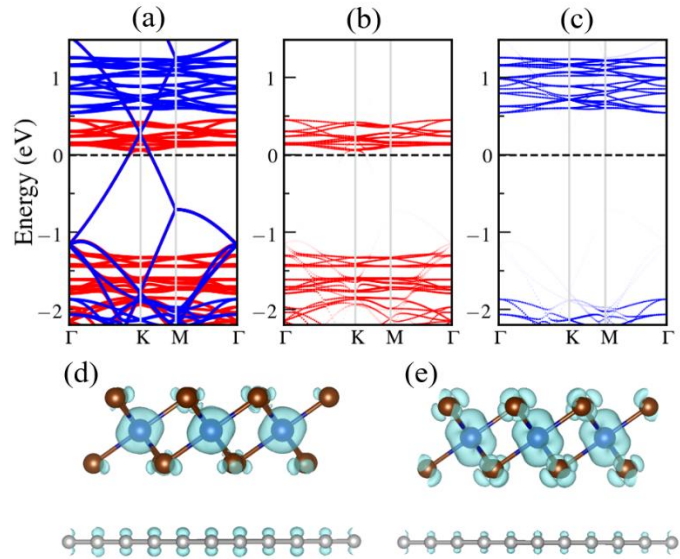

Figure 5 The spin-polarized band structure of CrBr3/Gr, (a), the red and blue curves denote spin-up and down band. The projected band of CrBr3 layer in CrBr3/Gr heterostructure in spin-up, (b), and down spin-down, (c), direction. The partial charge density of the bottom of conduction band in spin-up, (d), and spin-down, (e), direction. The isosurface is set to 0.0025 e/Å$^3$.

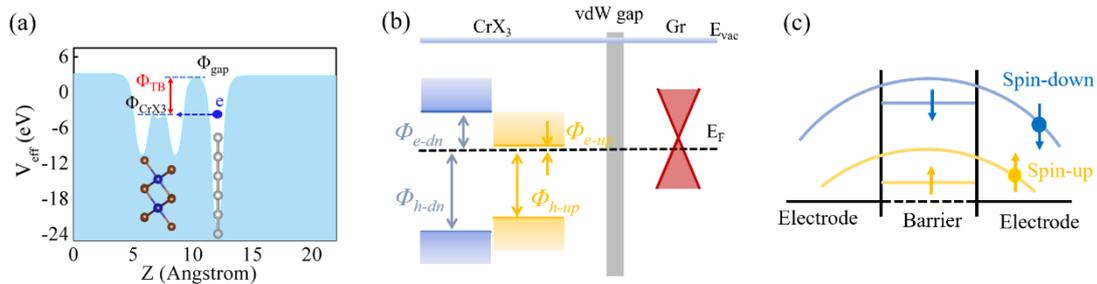

Figure 6 (a) Effective potential profile of CrBr3/Gr interface, (b) Schematic drawings of the band alignments for the CrX3/Gr heterostructures, (c) schematic energy diagram of barriers for spin-up and down electrons.

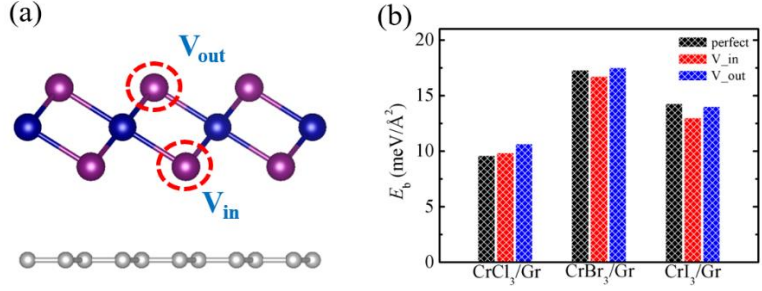

Figure 7 (a) Schematic drawing of two different configuration of defective heterostructures, (b) the binding energy E $_b$ for perfect, V*in* and V*out* CrX3/Gr heterostructures

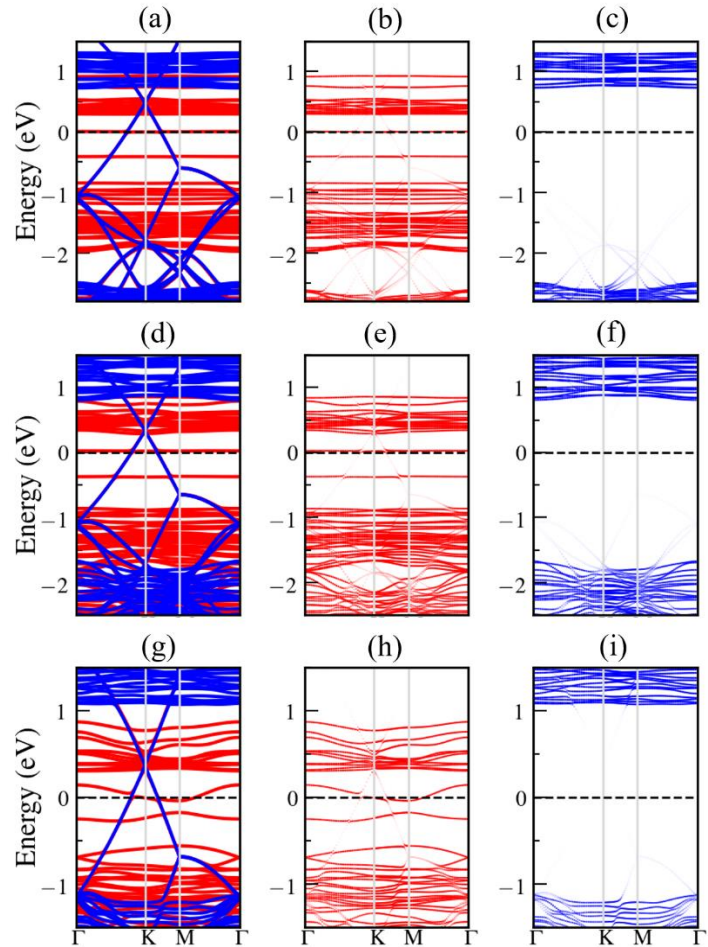

Figure 8 The spin-polarized band structure of defective (a) CrCl3/Gr, (b) CrBr3/Gr and (c) CrI3/Gr. The projected band of (b)(c) CrCl3, (e)(f) CrBr3 and (h)(i) CrI3 layer in spin-up and down direction in defective CrCl3/Gr, CrBr3/Gr and CrI3/Gr heterostructure, respectively. The red and blue curves denote spin-up and down band.

Table 1 The calculated parameters in CrX3/Gr. d, interlayer distance, in Å; $E_b$, binding energy, in meV/Å$^2$; $\Delta q$, transferred charge per CrX3 unit, in e.

|  | CrCl3/Gr | CrBr3/Gr | CrI3/Gr |
|---|---|---|---|
| d | 3.36 | 3.45 | 3.47 |
| $E_b$ | 3.36 | 17.38 | 14.34 |
| $\Delta q$ | 0.072 | 0.015 | 0.003 |

Table 2 The Schottky barriers for electron $\Phi e$ in spin-up, $\Phi e\_up$, and spin-down, $\Phi e\_dn$, in CrCl3/Gr, CrBr3/Gr and CrI3/Gr. The unit is meV.

|  | $\Phi_{e\_up}$ | $\Phi_{e\_dn}$ |
|---|---|---|
| CrCl3/Gr | 79 | 563 |
| CrBr3/Gr | 61 | 547 |
| CrI3/Gr | 34 | 759 |